\documentclass[runningheads,a4paper]{llncs}

\usepackage{url}
\urldef{\mailsa}\path|{alfred.hofmann, ursula.barth, ingrid.haas, frank.holzwarth,|
\urldef{\mailsb}\path|anna.kramer, leonie.kunz, christine.reiss, nicole.sator,|
\urldef{\mailsc}\path|erika.siebert-cole, peter.strasser, lncs}@springer.com|    
\newcommand{\keywords}[1]{\par\addvspace\baselineskip
\noindent\keywordname\enspace\ignorespaces#1}

\newif\ifpdf
\ifx\pdfoutput\undefined
	\pdffalse
\else
	\pdfoutput=1
	\pdftrue
\fi
\ifpdf
	\usepackage{graphicx,graphics,latexsym,verbatim,epsfig}
	\DeclareGraphicsExtensions{.pdf,.jpg,.png}
\else
	\usepackage{graphicx}
	\DeclareGraphicsExtensions{.eps}
\fi

\makeatletter
\def\maxwidth{
  \ifdim\Gin@nat@width>\linewidth 
    \linewidth
  \else
    \Gin@nat@width
  \fi
}
\makeatother

\usepackage{placeins}
\usepackage{amsmath}
\usepackage{amsfonts}
\usepackage{amssymb}
\usepackage{multirow}
\usepackage[noadjust]{cite}
\usepackage{fixltx2e}
\usepackage{array}

\begin{document}

\mainmatter  

\title{In the Dance Studio:  An Art and Engineering \\ Exploration of Human Flocking\thanks{This effort was supported in part by Princeton University's Essig Enright Fund, Lewis Center for the Arts, Keller Center for Innovation in Engineering Education,  and Mechanical and Aerospace Engineering Department, and by NSF  grant ECCS-1135724, AFOSR grant FA9550-07-1-0-0528 and ONR grant N00014-09-1-1074.}\thanks{Please cite as follows: Leonard N.E. et al. (2014) In the Dance Studio: An Art and Engineering Exploration of Human Flocking. In: LaViers A., Egerstedt M. (eds) Controls and Art. Springer, Cham, doi: \texttt{10.1007/978-3-319-03904-6\_2}.}}

\titlerunning{ An Art and Engineering Exploration of Human Flocking}

%
%
\author{Naomi E. Leonard%
\and George F. Young\and Kelsey Hochgraf\and Daniel T. Swain\and \\Aaron Trippe\and
Willa Chen\and Katherine Fitch\and Susan Marshall}  
\authorrunning{An Art and Engineering Exploration of Human Flocking}

\institute{Princeton University, Princeton, NJ 08544, USA}

%
%

\toctitle{In the Dance Studio}
\tocauthor{An Art and Engineering Exploration of Human Flocking}
\maketitle

\begin{abstract}
Flock Logic was developed as an art and engineering project to explore how the feedback laws used to model flocking translate when applied by dancers.  The artistic goal was to create choreographic tools that leverage multi-agent system dynamics with designed feedback and interaction.  The engineering goal was to provide insights and design principles for multi-agent systems, such as human crowds, animal groups and robotic networks, by examining what individual dancers do and what emerges at the group level.  We describe our methods to create dance and investigate collective motion.  We analyze video of an experiment in which dancers moved according to simple rules of cohesion and repulsion with their neighbors.   Using the prescribed interaction protocol and tracked trajectories, we estimate the time-varying graph that defines who is responding to whom.   We compute status of nodes in the graph and show the emergence of leaders.  We discuss results and further directions.  
\end{abstract}

\section{Introduction}\label{sec:intro}

The {\em Flock Logic} project \cite{flocklogic} was conceived by engineering professor/control theorist Naomi Leonard and dance professor/choreographer Susan Marshall.  The project was initiated as a joint exploration with professional dancers in July 2010 and then subsequently as a semester-long Princeton University Atelier course co-taught by Leonard and Marshall in Fall 2010. The professional dancers and the students, most of whom had previous dance training, participated in collaborative artistic and scientific investigations and experiments inspired by the complex and beautiful group motion that emerges in bird flocks and fish schools.  Leonard and Marshall's aim was to explore artistically and scientifically how individual rules of interaction and response within a network of dancers yield complex emergent collective motion of the group.  

The emergent nature of flocking and schooling was a central driver for the project: the remarkable collective motion of flocks and schools results not from a prescribed choreography nor even from a designated leader, but rather from simple rules of response that each individual obeys \cite{Parrish1999}.  These feedback rules govern how each individual moves in response to the relative position or motion of its close neighbors.   For instance, basic flocking rules typically have a cohesive element and a repulsive element \cite{Breder1954}.  The cohesive element requires that while each individual moves around it should remain a comfortable distance from a few others; the repulsive element requires that each individual should move away from others that get too close.   An active area of research is focused on explaining how the observed complex collective motion of animal groups emerges from, and is influenced by, the feedback rules, the dynamics of the social interactions within the group, the distribution of information across the group, the features in the spatial surrounding, the differences among individuals, the noise in measurements, and the uncertainty in decision-making \cite{CouzinSurvey2003,sumpter}.   Analytical and numerical studies and laboratory and field experiments have all been used to investigate; for example, see \cite{star1,Hemelrijk2010,YoungStarlings2013} for a range of studies on flocking of starlings.

Flock Logic explored what happens when a group of {\em dancers} apply these and related feedback laws as they move around a space together.  In the Flock Logic explorations the flocking rules were prescribed, but neither how the dancers applied the rules nor how faithfully they followed the rules were  controlled.  For example, the number of neighbors and the distance from neighbors to maintain in the coherence rule were prescribed, but the dancers were not instructed how to choose with whom to cohere, how to prioritize among neighbors moving in diverging directions, nor how to handle  conflicts such as when cohering with one dancer meant getting too close to another dancer.  It was also possible that dancers broke the rules at times.   Thus, the emergent human flocking resulted from both prescribed and individualized, and thus unknown, features of dancers' choices and dynamics.   In this way the Flock Logic project provided a framework for exploring emergent collective behavior  somewhere between studying animal aggregations in the wild, with all that is unknown, and examining computer simulated flocking, with its exclusive reliance on a prescription.    

This aspect of Flock Logic made it particularly well suited to an integrated art, engineering and science agenda.   On the one hand, the Flock Logic framework made it possible to observe the influence on  collective motion of natural biases, in this case human biases, and heterogeneity across the group.  Dancers with different physical features, personalities, dance training, etc., would respond differently  to one another and would prioritize rules and resolve conflicts differently.    This would affect how information would pass through the group and how the group as a whole would respond to external forces.    On the other hand, the Flock Logic framework made it possible to systematically examine the influence on collective motion of parameters of the prescribed rules.  This applied to the parameters of  rules and  environments meant to represent  animal groups, e.g., number of others with whom to cohere,  total population of the group, availability of information or preferences across the group, as well as the shape, size and placement of obstacles.   This also applied to the parameters of  rules and environments not necessarily intended to represent animal groups but rather motivated by artistic  and engineering design goals.  By varying rules and environmental features beyond what one would expect in animal groups, it was possible to explore how individual-level behaviors connect more generally to the  aesthetics and the functionality of the emergent group-level behaviors.  And this led to the creation of a wide range of artistic and engineering design possibilities.

Leonard's engineering goal was to use the explorations with dancers to gain insight into the mechanisms of animal group and human crowd dynamics and into design principles for control of natural and robotic groups.  Could the dance studio be viewed as an experimental test-bed in this regard?   Could the human data collected be used to help explain a range of collective behaviors?  The dancers could represent a human crowd moving  in a bounded space, trying to avoid colliding.  Likewise, dancers moving in a studio, responding to local neighbors and the environment, provide a reasonable approximation to the collective motion of a herd. The walls of the studio are like trees or topography, and the heterogeneity among the dancers (experience, height, confidence) is similar to that in a herd \cite{Gueron1996}.   Further, dancers are particularly well suited to these kinds of explorations because they are trained to be physically aware and can comfortably handle a number of feedback rules.   
Thus, the setting provided enormous flexibility in the kinds of questions that could be addressed.  For example, in the present chapter, the human motion data are used to rigorously study how influence among individuals in the network is distributed and how that is reflected in the changing spatial distribution of individuals and in the group-level shape and motion dynamics.   This could, for example, lead to insights on how human crowds move in cluttered spaces and how animals organize themselves to reduce vulnerability to predators \cite{Rubenstein2007}.  This could also lead to bio-inspired methods for designing robust and responsive networks of heterogeneous robots \cite{Nolcos13}.
  
There are a number of motivating and complementary scientific studies of human collective motion, many of which focus on crowd dynamics.  Experiments on leadership and decision-making in human crowds were described in \cite{Dyer2009}.  In \cite{Theraulaz2010} analysis of natural pedestrian group motion revealed  the influence of social interactions on crowd dynamics.  In \cite{Bonabeau2004} a design method for 
human collective behaviors used evolutionary dynamics.   Simplified models described in  \cite{moshpit} predicted collective behavior of humans in mosh and circle pits as observed from video data of heavy metal concerts.

Leonard's engineering goal was tightly integrated with Marshall's artistic goal.  Marshall approached Leonard in 2010 about the possibility of an investigation and a course to find out what would happen if a group of dancers were in possession of the rules governing the motions of groups of flocking animals. Marshall could see how applying Leonard's work on decentralized control of collective motion, to dancers, could potentially result in choreographic tools or training tools for developing individual and group awareness.  She also imagined a site-specific large group performance work developed with little more than a site and the rules. 

Marshall's interest centered on a desire to translate flocking rules such as those related to group cohesion and response to external pressures into improvisational instructions for dancers.  Could these rules support unexpected and complexly orchestrated collective motion to emerge from individual interactions? How might the local sensing rules be altered choreographically to make emergent choreography that didn't resemble the familiar look of organic flocking? Could these rules be learned quickly by non-dancers to create a kind of flash mob performance? 

In theater and dance, there is a long history of movement practice and performance based on structured improvisation and rules and games  \cite{Clemente1990}.  Often in improvisational dance work, the individual has a wide range of choices open to them and takes compositional responsibility for the entire stage as well as their own body.  The flocking rules tend to limit the individual's choices to their immediate neighbors and to ask the individual to relinquish group choreographic responsibility -- nonetheless, rich group choreography results. Could rules be designed that would allow dancers, ignorant of any overarching choreographic goals, to create complex and organized patterns using these tools? 

To generate human flocking, the dancers were asked to move around a space and follow rules that were defined in advance.   To enable {\em cohesion}, each dancer was given the rule to keep $m$ of their neighbors at a distance of arm's length with the selection of the $m$ neighbors freely changeable.   To enable {\em repulsion}, each dancer was asked to avoid letting any dancer get closer than arm's length.    To prevent tripping, the dancers were asked to avoid moving backwards.   

These three rules (cohesion, repulsion, backwards avoidance) were among the most fundamental rules examined.   Variations on the three fundamental rules were prescribed as well as a range of additional and alternative rules.   For example, rules for alignment with neighbors, response to obstacles and walls, options to initiate or imitate specific movements, etc., were implemented.   More complex informational structures were imposed -- for example, two or three dancers in the group were secretly given additional rules, such as to move to a particular location or according to a particular pattern.   The dancers also performed rules for other kinds of behaviors such as dynamic coverage and pursuit and evasion. In part because each dancer's motion was relatively under-prescribed, there was considerable room for variation among individuals, e.g., in speed, facing direction relative to motion, selection of neighbors, positioning relative to neighbors, and response to walls or obstacles.  

\begin{figure*}
\centering
\includegraphics[width=0.95\textwidth]{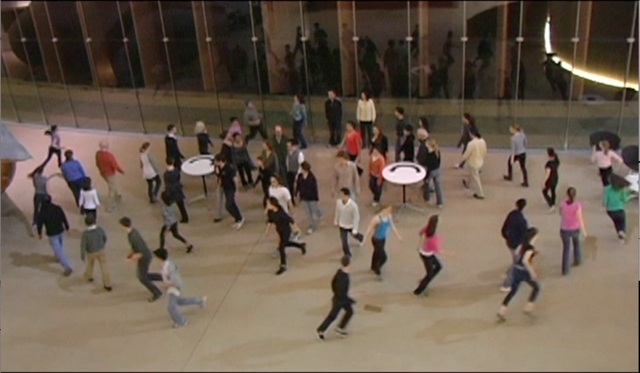}
\caption{Snapshot from a Flock Logic performance at Princeton University in December 2010. }
\label{fig:perf}
\end{figure*}

Complex and artistically satisfying collective behaviors were routinely observed. As  part of the 2010 Princeton University Atelier course, approximately fifty volunteers participated in two flocking performances, each at a different site, after having been briefly instructed in a few local rules of cohesion, repulsion and alignment as well as responses to obstacles, to walls, and to ``predators''. From Leonard and Marshall's perspectives, these were highly satisfying performances.   A snapshot from one of the performances is shown in Figure~\ref{fig:perf}.   Video clips from the events are publicly available and can be accessed from the Flock Logic website \cite{flocklogic} or directly at the following links: \\  {http://vimeo.com/19361231} (Peter Richards);  \\
{http://www.princeton.edu/main/news/archive/S29/62/38K80/} (Evelyn Tu);  \\
{http://www.youtube.com/watch?v=Mg29hawdcMw} (Jeffrey Kuperman).

In this chapter, we describe the Flock Logic explorations and the tools used for our artistic and engineering investigations.   As an illustration, we examine one experiment with thirteen dancers who followed the flocking rules of cohesion and repulsion as they moved around the dance studio.   Using the trajectories tracked from an overhead video camera and the prescribed interaction rules, we estimate the time-varying graph that encodes who is sensing whom as a function of time.    We compute the time-varying status of each node in the graph, defining how much attention a dancer receives from the rest of the dancers, and use these to infer emergent leaders.   We discuss implications, open questions, and further directions both artistic and scientific.
  
In Section~\ref{experiment} we  describe our human flocking explorations, including our on-line FlockMaker software tool, and the human flocking experiments. Trajectory tracking is described in Section~\ref{tracking}.  In Section~\ref{graphs} we review graphs and FlockGrapher, our  tool for visualizing graphs.   In Section~\ref{model}, we estimate the time-varying graph of the network.   In Section~\ref{analysis} we estimate node status  and discuss the influence of individuals.  We conclude with a discussion of the results and a reflection on further artistic and engineering opportunities that build on the Flock Logic project in Section~\ref{final}.

An earlier version of this paper appeared in the proceedings of the American Control Conference, held in Montreal, Canada in June 2012 \cite{FLACC}.   At this  same conference, a special interactive session was held in which 100 conference participants participated in a human flocking performance event.

\section{Human Flocking} \label{experiment}

\subsection{Explorations}

A typical Flock Logic exploration involved on the order of 10 to 15 dancers who moved around the dance studio for a few minutes applying flocking rules prescribed by Leonard and Marshall.   Many of these explorations were run in series during a single session, with a wide variety of flocking rules prescribed.   The dancers were given frequent opportunities to watch the group from the outside, and to discuss how it felt from within the group and how it looked from without the group.   The process was highly collaborative:  dancers made suggestions routinely and during a number of sessions small groups  of dancers  would design a set of rules for themselves and for the rest of the group.      

The explorations evolved over time as the dancers gained more experience with moving according to the rules of flocking.   This meant that level of experience played into the emergent collective motion, especially later in the semester when volunteers were briefly ``trained" and joined the group for flocking.   By adding dancers to the group, the role of the number of dancers was also explored.   As many as 24 dancers participated in some of the experiments in the studio, described in Section~\ref{experiments}.   In one of the Flock Logic performances in December 2010  more than 50 people were involved and in later flocking events, such as at the special session in Montreal at the American Control Conference, as many as 100 people participated.     Sessions were also held outdoors which provided the opportunity for explorations in a space without boundaries.  

The basic flocking rules of cohesion, which meant keeping $m$ neighbors at arm's length, repulsion, which meant moving away from anyone closer than arm's length, and backwards avoidance, formed the basis of many explorations.   Two-person cohesion ($m=2$) was enough to create what looked like a planar school or flock.   With $m=2$ the dancers were regularly spaced and exhibited polarized motion, i.e., everyone moving together in a single direction, as well as circular motion, i.e., the group moving around a circle.  The circular motion sometimes drifted and sometimes remained fixed about a single stationary dancer.   The group also experienced fissions and fusions as well as significant changes in momentum.   When $m$ was decreased to $m=1$, i.e., one-person cohesion, the result was a lot more ``parading'' and much less of the distributed look of a natural flock.   

Explorations made use of the presence of walls and the introduction of obstacles.   With no special rules attached to walls or obstacles, the dancers treated them much as animals would likely treat them, i.e., they deftly avoided colliding with them.    In one case a row of chairs was extended in a line from one wall into the middle of the room.   When the dancers moved into the smaller space created by the chairs and the parallel wall, they would remain there temporarily, as if caught in a tidal pool, and only move out once they had reversed direction.   When rules were prescribed with respect to walls and obstacles, all sorts of interesting, and less biologically motivated, collective behavior emerged.   For example, the walls were given an attractive pull as well as a ``stickiness''.   As a result, dancers getting close to a wall got pulled away from the group and stuck there (see Figure~\ref{fig:perf}).   They were released from the wall by cohesion to other dancers when at some later point the group passed by them.    This looked like the peeling off and adhering back of dancers in an ordered way since these dancers still applied the cohesion and repulsion rules among themselves.   

In another case, a  round table was moved into the open space and endowed with the properties of a sling shot.   When dancers got close to the table they would circle around it in a fixed direction, e.g., counter-clockwise, at an increased speed and then get ``flung off.''   This led to a variety of different outcomes since different dancers resolved conflicts differently.   For example, in early runs with the round table, the first few dancers moved around the table at elevated speed, but then they tended to slow down and congregate in a slow moving flock nearby.   The dancers who went around the table subsequently either stopped short to avoid cutting through this congregating group or broke the repulsion rule and charged right through it.  In the performance events in December 2010,  a ``waterfall'' effect was sustained with two slingshot tables (see Figure~\ref{fig:perf}). 

The ability to prescribe individualized preferences and objectives was also explored.  These explorations were motivated by an interest in understanding the role of heterogeneity in preference and objective in groups,  if and how individuals can exert leadership through motion, and the range of emergent collective motion patterns that can result. Individualized objectives were typically prescribed secretly: all dancers were told to follow the three basic flocking rules and 2 or 3 of the dancers, unbeknownst to the others, were directed to  follow certain additional  rules.   For example, the 2 or 3 selected dancers were sometimes given the same additional rule, such as to head for one corner of the room or out a door.  Alternatively, the 2 or 3 dancers were given conflicting rules, such as one told to aim for one corner of the room and another to aim for the opposite corner.  Or the 2 or 3 dancers were given a joint objective such as to split the group into subgroups.   We explored how dancers attempted to attain their additional objectives, and we observed and discussed how some dancers were successful and some were not successful in influencing the other dancers through their motion.   

\begin{figure*}
\centering
\includegraphics[width=1\textwidth]{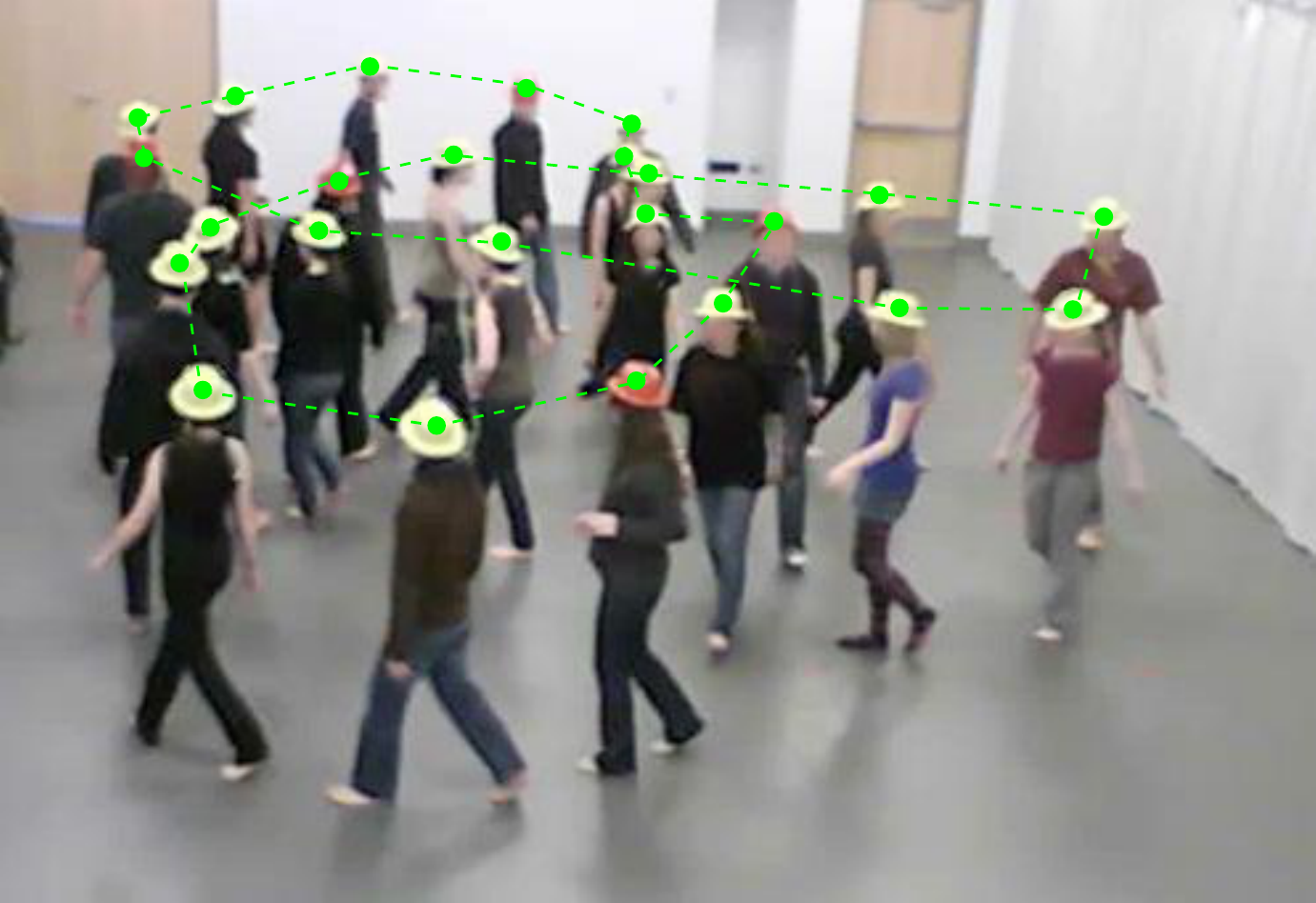}
\caption{Snapshot of a cyclic pursuit experiment with 24 dancers.   The position of each dancer is marked with a green dot and a dashed line connects each pair of pursuer and pursued.   The collection of dashed lines shows that the dancers move around a single closed curve that loops around, intersecting itself three times. }
\label{fig:cycpursuit}
\end{figure*}

In some explorations, 1 or 2 dancers were instructed to behave as a predator or pursuer by waving a hand or t-shirt or flashing a bicycle light.   In this case the dancers were given the rule to keep a safe distance from the pursuers, e.g., 5 or 10 feet depending on the size of the studio.   The pursuers could thus put pressure on the group and create a variety of beautiful patterns by trapping the flock, shaping the flock, and restricting the flock's motion to changing corridors of space in the studio.     Cyclic pursuit and evasion was also explored in which case every dancer was assigned one other dancer to pursue so that the group made a closed cycle with each dancer having one person to pursue and one person to evade.    The motion patterns were constantly changing loops with multiple intersections, with qualitative features as predicted in \cite{pais2010}.    Figure~\ref{fig:cycpursuit} shows a snapshot of 24 dancers in a cyclic pursuit experiment in the dance studio.   

We found that adding an optional alignment rule created further artistic variation and options.   We also explored many other rules in place of the basic flocking rules; these contributed significantly to artistic, engineering and scientific goals.   For example  a rule was applied in which each dancer moved with oscillating speed, i.e., accelerating and decelerating repetitively,  and such that the oscillations were out of phase with others nearby.  This was motivated by the oscillating speed observed in fish schools and the rich family of motion patterns that could be designed using this rule  \cite{swain}.   In another example, rules for dynamic coverage were explored; these rules were motivated by problems of foraging over spaces of distributed resource \cite{carlos}.
 
\subsection{FlockMaker}

{\em FlockMaker} is a Java WebStart application developed to aid the Flock Logic project and  designed for simulation and exploration of  collective motion \cite{flockmaker}.  FlockMaker is intuitive for a curious layperson, and  it can be used to model complex combinations of flocking rules and initial configurations. 
In the model, each dancer is represented as a single particle moving in the horizontal plane with variable velocity. Speed and facing angle (but not acceleration) are taken to be approximately continuous in time. 

The simulation user can assign a variety of flocking rules to the dancers, such as ``Pursue Someone," ``Repel Neighbors," and ``Slow Down Near Neighbors." To further control behavior, the user can set values for a wide range of parameters pertaining to a dancer's rules or initial configuration, including radius of sensing, number of neighbors sensed, maximum speed of rotation, and magnitude of additive random noise. 
Different rules can be assigned to different dancers. Furthermore, each dancer can be assigned to follow multiple rules at a time, each rule potentially carrying a different relative weight representing its level of priority. 

Dancers interact not only with each other, but also with the room in which they are moving, represented as a rectangular space contained within four walls. The FlockMaker user can change the size of the room, add obstacles to the room, and add rules applicable only within certain zones of the room. 

After several weeks of work in the studio, the students in the Princeton University Atelier course spent time using FlockMaker, both to test ideas that had been tried in the studio and to investigate new ideas.   Several of the rule sets and emergent behaviors investigated in FlockMaker were subsequently explored in the studio.

\subsection{Experiments}
\label{experiments}

A series of human flocking experiments was run in mid December 2010 in the 62' 7" x 28' 4" New South dance studio at Princeton University.   Groups of dancers carried out the three basic rules of flocking with manipulations on initial conditions, number of dancers $N$ (either 13 or 24), and number of neighbors $m$ for cohesion (either 1 or 2).  Alignment with neighbors was tested as was the assignment of an additional rule to two of the dancers (of which the others were not aware), which was to try to split the group.   Several experiments were also run with dancers implementing the rules for cyclic pursuit; see Figure~\ref{fig:cycpursuit}.   

Six Trendnet IP-600 cameras, synchronized over a local wired network, were set up in fixed locations to record the motion of the dancers.  Two cameras were hung on the ceiling near either end of the studio, facing inward towards each other, and  four were mounted high up on one side wall.  Camera views covered the majority of the space in the studio and overlapped significantly.  Using built-in software, the cameras recorded video and stored it on a laptop.   The video provided 640 x 480 resolution and 20 frames per second.

For the December 2010 series of experiments, part of the room was blocked off so that the motion of all of the dancers could be fully captured by one of the six cameras (one of the two fixed to the ceiling).   The dancers wore bright colored hats, black clothing and bare feet to aid trajectory tracking.  

In this chapter we examine one experiment from the series in which there were $N=13$ dancers --  two professional dancers and eleven students.  
All thirteen dancers were asked only to follow the three basic rules of flocking with cohesion to $m=2$ neighbors.   The total time for  the experiment was 185 seconds, corresponding to the period from the start to the stop of the music.   We study the tracked trajectories of the dancers from the first 72 seconds of this experiment.

\section{Trajectory Tracking} \label{tracking}

\begin{figure}
\centering
\includegraphics[width=\maxwidth]{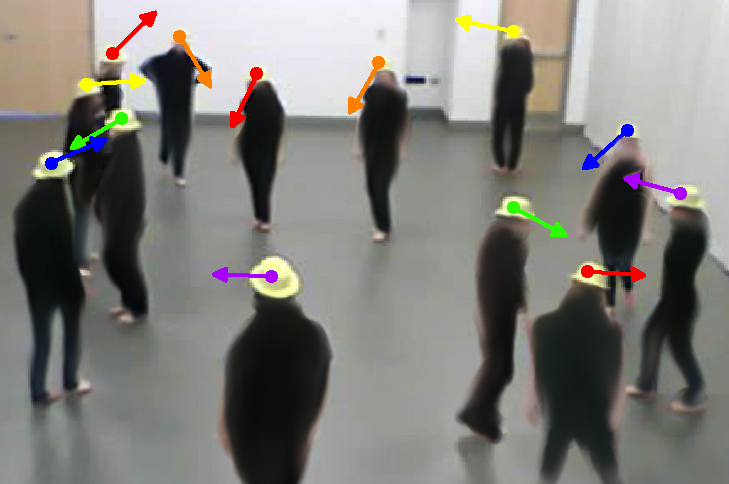}
\caption{From \cite{FLACC}.  One frame from the video of the experiment with superimposed tracked positions (colored dots) and normalized velocity vectors in the image plane (colored arrows indicating the directions in which the centroids of each hat are moving in the image).  Images of dancers are deliberately blurred.}
\label{fig:trackedframe}
\end{figure}

Trajectories were estimated using custom tracking software applied to the overhead video from one camera for the first 72 seconds 
of the experiment.  The tracked trajectories comprise an ordered  set of 1440 planar position vectors $(x,y)$  for each of the thirteen dancers.   A velocity vector is computed for each dancer at every time step by differencing the position vectors.   Speed and heading are computed as the magnitude and angle of the velocity vector.   Figure~\ref{fig:trackedframe} shows one frame from the video with superimposed  tracked positions and directions of motion. 

The custom tracking software uses a modified version of a real-time tracking algorithm that we have developed and used successfully for experiments involving multiple fish and robots \cite{SwaCouLeo2011}.  The algorithm is implemented using the MADTraC C++ library~\cite{MADTraC}, which in turn relies upon OpenCV~\cite{OpenCV} for low-level image processing routines.  The original tracking software was designed to address the challenges of tracking potentially densely distributed objects that are very similar to one another in appearance.  It was therefore applicable to the task of estimating dancers' trajectories.   

The tracking algorithm follows three steps that are iterated for each video frame, and described in greater detail in \cite{SwaCouLeo2011}.  In the first step, image segmentation produces a set of ``blobs'', such that each blob is a collection of contiguous pixels with high likelihood of belonging to any dancer's hat.   Likelihood is determined by thresholding each pixel's value in HSV color space and mapping to a binary image.  Blobs are extracted from the binary image using OpenCV's built-in blob labelling algorithm, which is based on \cite{Chang04}.  A blob is often associated with more than one dancer because of the physical proximity of dancers to one another, the proximity of dancers in the image due to the  angle of the camera, and noise in the image.  

In the second step, the blobs are analyzed in order to extract a noisy measurement for the position of each dancer.   If a single dancer is associated with a blob, then the measurement of that dancer's position  is taken as the centroid of all pixels  in that blob.  Otherwise, to resolve multi-dancer blobs or clusters of densely-spaced blobs, an expectation-maximization mixture-of-gaussian (EMMG) algorithm is used, which iteratively adjusts dancer positions for a given cluster and provides  position measurements as output. 

In the third step, the noisy position measurements are used  with an unscented Kalman filter (UKF) for each dancer to provide a more accurate estimated position $(x,y)$ in the current frame and to predict the position in the next frame.  The estimated position of each dancer is stored as the current point in the dancer's tracked trajectory.  The predicted positions are used to inform the next tracking iteration.  The $(x,y)$ position vector is expressed in a coordinate frame that is parallel to the floor.  The transformation to these coordinates from image plane coordinates was determined by applying camera calibration techniques to an image of several objects placed at known locations in the scene.  The average height of each dancer is assumed to be 1.65 meters.

\section{Graph Theory and Visualization} \label{graphs}

\subsection{Background on Graphs} \label{background}

Let $N$ be the number of  dancers.   For each dancer $i$ we define the set of neighbors, $\mathcal{N}_i$, to be the set of dancers whose positions are observed and used for cohesion by dancer $i$.

We associate to the system a \emph{sensing graph} $\mathcal{G} = \left( \mathcal{V}, \mathcal{E}, A \right)$, where $\mathcal{V} = \left\{ 1, 2, \ldots, N \right\}$ is the set of nodes, $\mathcal{E} \subseteq \mathcal{V}\times\mathcal{V}$ is the set of edges and $A$ is the $N \times N$ adjacency matrix with $a_{i,j} = 1$ when edge $\left(i,j\right) \in \mathcal{E}$ and $a_{i,j} = 0$ otherwise. Every node in the graph corresponds to a dancer, and the graph contains edge $\left(i,j\right)$ when $j \in \mathcal{N}_i$. An edge $(i,j) \in \mathcal{E}$ is said to be \emph{undirected} if $(j,i)$ is also in $\mathcal{E}$; otherwise it is \emph{directed}. A graph is undirected if every edge is undirected, that is, if $A$ is symmetric; otherwise it is directed.

A graph can be represented visually by drawing a dot for each node and a line between the appropriate pair of nodes for each edge. An undirected edge is typically drawn as a simple line, while a directed edge $\left(i,j\right)$ will have an arrow head pointing from node $i$ to node $j$.

A \emph{path} in $\mathcal{G}$ is a (finite) sequence of nodes containing no repetitions and such that each node is a neighbor of the previous one. The length of a path is given by the number of edges traversed by the path. The \emph{distance}, $d_{i,j}$, between nodes $i$ and $j$ in a graph is the shortest length of any path from $i$ to $j$. If no such path exists, $d_{i,j}$ is infinite.   This distance is not a metric since $d_{i,j}$ is not necessarily equal to $d_{j,i}$.  

The graph $\mathcal{G}$ is \emph{connected} if it contains a globally reachable node $k$; i.e. there is a path in $\mathcal{G}$ from $i$ to $k$ for every node $i$.  $\mathcal{G}$ is said to be \emph{strongly connected} if there is a path between every pair of nodes in the graph. A strongly connected component of $\mathcal{G}$ is a maximal subset of nodes such that there is a path in $\mathcal{G}$ between every pair of nodes in the subset. $\mathcal{G}$ is \emph{weakly connected} if it is connected when every directed edge is replaced by an undirected edge. A weakly connected component is a maximal subset of nodes that forms a  connected component when every directed edge in $\mathcal{G}$ is replaced by an undirected edge.

The \emph{status}, $s_k$, of a node $k$ is the average inverse distance between every other node and $k$. That is, $s_k = \frac{1}{N-1}\sum_{j \neq k}{\frac{1}{d_{j,k}}}$. $s_k$ will be maximum (equal to $1$) if there is an edge from every other node to node $k$, and minimum (equal to $0$) if there are no edges leading to node $k$.

\subsection{Visualization of Graphs} \label{visualize}

FlockGrapher is our Matlab tool that computes, visualizes and evaluates different kinds of graphs derived from flock position data. Using a graphical user interface, the tool accepts  tracked position and direction of motion data for individuals in a flock in two or three dimensions.  It can visualize data from one specific instant in time or create a time series animation of  data sets corresponding to successive time steps. The user can create graphs from the data by defining an individual's neighborhood in terms of  either  a prescribed number of nearest neighbors or  a prescribed sensing radius. For data that includes the direction of motion of nodes, FlockGrapher can incorporate a limited viewing angle, 
assumed to be symmetric about the individual's direction of motion.  In the case of a fixed number of nearest neighbors and a limited viewing angle, if there are fewer than the required number of neighbors  visible to a node,  the  viewing angle will be rotated with respect to the direction of motion until enough neighbors are visible. Edge weights can be automatically manipulated, e.g., as a function of distance between nodes, or they can be prescribed by the user.  

\begin{figure*}
\centering
\includegraphics[width=1\textwidth]{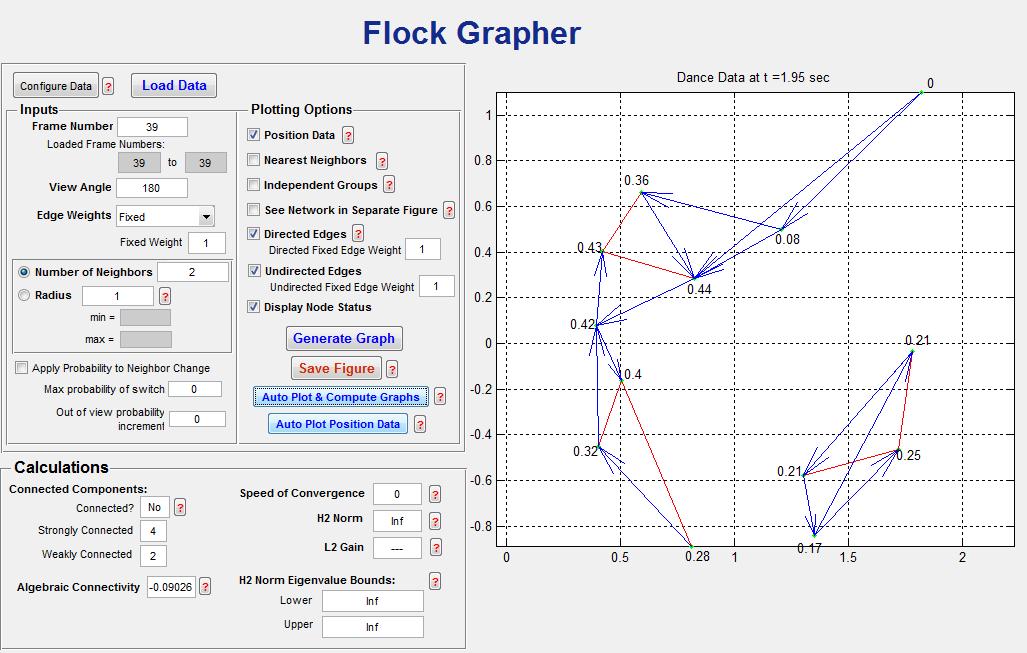}
\caption{Screenshot of FlockGrapher using dancer data corresponding to the instant shown in Figure \ref{fig:trackedframe}. Nodes are shown as small green circles connected by edges.  Directed edges are  blue with arrow heads and undirected edges are  red.  The number next to each node is its   status.  Other computed graph properties are displayed on the left.}
\label{fig:flockgraph}
\end{figure*}

Once a sensing graph has been computed, FlockGrapher can evaluate a range of graph properties, including   number of strongly and weakly connected components,  algebraic connectivity, speed of convergence and node status.  
The tool also displays some  properties on the graph visualization; for example, directed and undirected edges can be distinguished with different colors.  For sets of data corresponding to successive time steps, the time-varying values of these properties will be displayed as the graph visualization changes.  In the case of the human flocking experiment, this dynamic graph visualization can be run at the same time as the video of the dancers to compare computed and observed behavior.   FlockGrapher can save all the computed data to allow for further analysis. A screenshot of FlockGrapher is shown in Figure \ref{fig:flockgraph}; the graph and its properties corresponds to the frame from the video shown in Figure~\ref{fig:trackedframe}. 

\section{Sensing Model and Graph Computation} \label{model}

Since each dancer was given the same specific rules to follow, it is in principle possible to apply the same rules to our tracked data and reconstruct the sensing graph used by the dancers. However, certain aspects of both the rules and human behavior make this task challenging. Although the dancers were each told to stay arm's length from two other dancers, no instruction was given for how they were to choose these two neighbors. In addition, although humans have a field of view of up to $200^\circ$ \cite{Werner1991}, there was no compulsion for the dancers to keep both of their neighbors visible at all times.

Given these limitations, we make two key assumptions in order to estimate the dancers' sensing graph. First, we assume that each dancer only chose neighbors from within a limited angular range centered about the direction they were travelling.   Since no dancer was observed to be rapidly moving their head, the direction of motion is assumed to be a reasonable proxy for direction of the head and therefore for center of viewing range. Although this assumption is generally applicable to the data, there were instances observed in which a dancer would either move in a different direction to where they were facing, or move with their head turned at a constant angle to their body. These occurrences are impossible to detect with our point-tracking approach, but could be accounted for with a more sophisticated tracker with the ability to detect the orientation of each dancer's face. Second, we begin by assuming that each dancer was applying the cohesion rule with the two {\em nearest} neighbors within this range. Since every dancer was trying to keep  two neighbors at arm's length (and let no dancers closer than arm's length), a dancer's neighbors would naturally be among the closest of the other dancers. 

With these assumptions we used FlockGrapher to estimate the sensing graph at each time (frame) by computing the two-nearest neighbor graph with a limited viewing angle. When fewer than two other dancers were visible using the direction of motion to center the viewing region, this region was allowed to rotate until two dancers became visible. However, we did not know \emph{a priori} what viewing angle to choose to best represent the dancers' behavior.

For collective behavior, it is impossible to guarantee that a group will remain together if the communication graph is not connected \cite{Ren2005}. When the graph is disconnected, there is nothing to prevent different subgroups from moving in different directions and splitting the group. However, other features of the environment (such as the limited space in the room) can drive the group back together.  Since fissions and fusions of the group were observed, we selected  the viewing angle for our sensing model as the one that produced a graph that  was disconnected when the group of dancers was observed to split and remained connected when the group of dancers was observed to be  cohesive. 

Table \ref{tab:viewangle} shows the results of estimating the sensing graph across the whole tracked period using different viewing angles. We can see that reducing the viewing angle from $360^\circ$ to $270^\circ$ significantly improves the amount of time the graph is connected, with the maximum connectedness occurring with a viewing angle of $120^\circ$. However, our goal was not simply to maximize connectedness but rather to match the observed behavior of the group.

Early in the experiment, between about $1$ and $3$ seconds, a small group of four dancers split from the rest of the group. The dancers within this group appeared to be observing only one another. Eight of the remaining dancers also formed a group, only observing one another. The final dancer was  originally able to observe both groups before turning to face the larger group, but since no other dancer was observing this individual, the group was split during this whole period. Eventually, the dancers in the larger group turned and observed the smaller group, leading to a single ``flock'' again. This disconnection event was reflected in the estimated graphs for viewing angles of $150^\circ$ and greater, but not for the smaller angles. However, with a viewing angle of $150^\circ$ the graph became connected at a few points within this interval when direct observation of the video suggests that the group was still split. This was not the case with a viewing angle of $180^\circ$; thus, $180^\circ$ was chosen as providing the best match of the splitting behavior of the dancers. Figure \ref{fig:trackedframe} shows the group during this disconnection event and the graph in Figure \ref{fig:flockgraph} (corresponding to the  frame of Figure~\ref{fig:trackedframe}) is computed using a viewing angle of $180^\circ$.

\begin{table}[thb]
\centering
\caption{Effects of viewing angle on graph connectedness over the  whole tracked period}
\label{tab:viewangle}
\begin{tabular}{| >{\centering} m{0.2\linewidth} | >{\centering} m{0.3\linewidth} | >{\centering\arraybackslash} m{0.3\linewidth} |}
\hline
Total viewing angle & Percentage of time connected & Number of disconnection events \\
\hline
\hline
$360^\circ$ & $59.58\%$ & $40$ \\
\hline
$270^\circ$ & $91.67\%$ & $43$ \\
\hline
$210^\circ$ & $97.5\%$ & $10$ \\
\hline
$180^\circ$ & $98.47\%$ & $3$ \\
\hline
$150^\circ$ & $98.68\%$ & $5$ \\
\hline
$120^\circ$ & $99.65\%$ & $3$ \\
\hline
$90^\circ$ & $99.58\%$ & $3$ \\
\hline
\end{tabular}
\end{table}

Although our first estimate of the sensing graph captures a split in the group and stays connected during the rest of the tracked period, it remains a crude approximation to the true sensing graph. For example, some nodes change their neighbors rapidly in our estimated graph, which is likely an overestimation of the rate at which dancers switch neighbors. Instead, if an individual has just been chosen as a neighbor, that individual is likely to stay a neighbor for a period of time rather than being immediately discarded as another individual comes closer in view. Two steps were taken in an effort to reduce rapid neighbor fluctuations. First, the tracked position data was passed through a low-pass filter, which consequently smoothed out node headings. Then, to account for an individual's reluctance to change neighbors soon after they are chosen, we added a term to the estimation model representing the probability of switching from a current neighbor to a closer dancer.  The lower was the probability the greater was the ``inertia'' of the dancer to switch to a closer dancer, equivalently, the greater was the commitment of the dancer to its current neighbor.   By allowing this probability to reset to a low value whenever a new neighbor was chosen and then increase with time, we could capture the inertia of edges in the sensing graph.

\section{Analysis of Individual Influence} \label{analysis}

\begin{figure}
\centering
\includegraphics[width=\maxwidth]{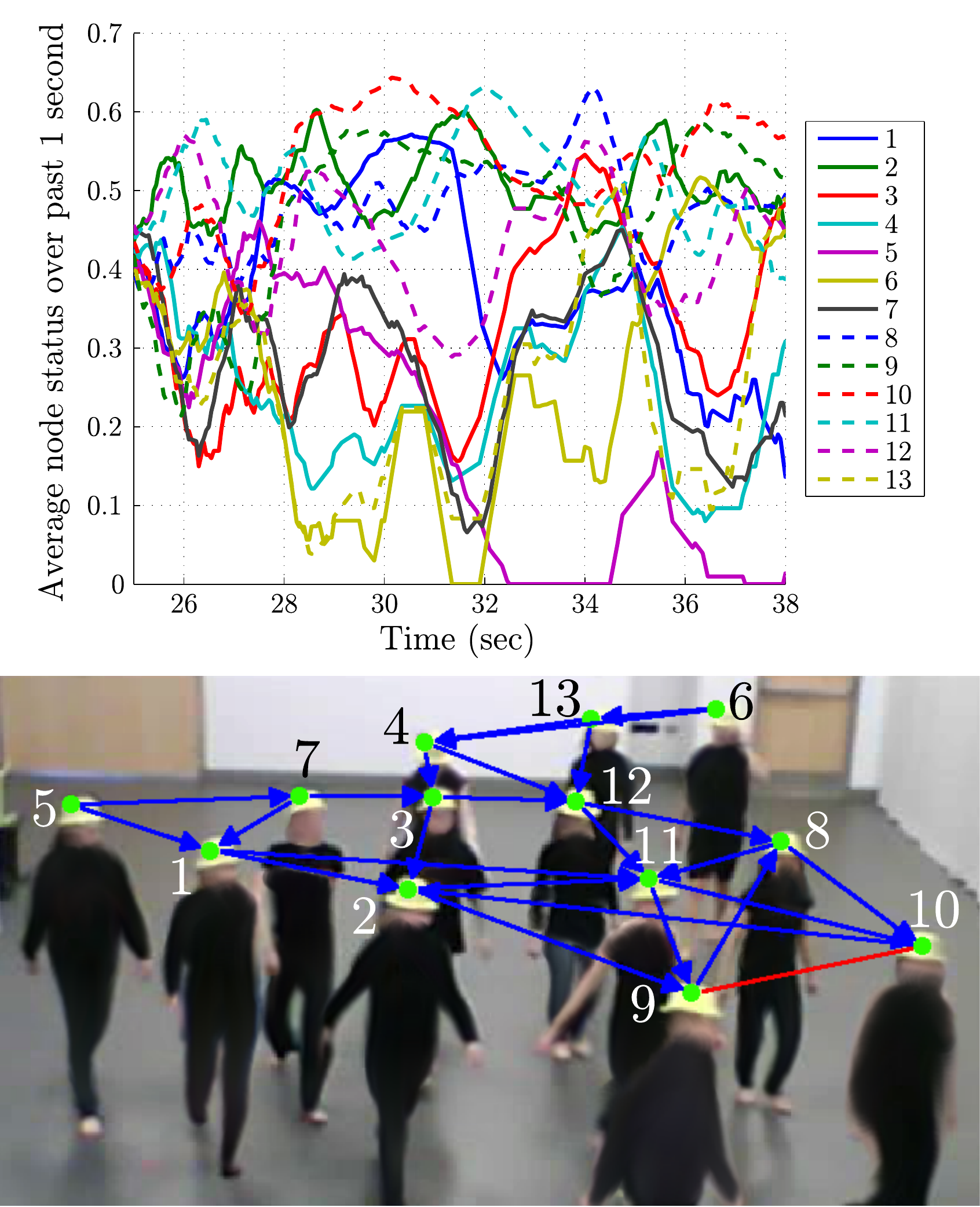}
\caption{From \cite{FLACC}.  Plot of $1$-second running average of node status, along with a sample video frame and sensing graph near the end of the leadership event from $t = 28.75$ seconds to $t = 31.45$ seconds. The red edge is undirected while all blue edges are directed.  We can observe that node 10, with the highest status, corresponds to the dancer in the front of the group.}
\label{fig:status}
\end{figure}

We use the estimated time-varying sensing graph to begin investigating the influence of each individual within the group. Our method is to compute and compare node status. Without knowing precisely how each individual implemented the flocking rules, node status can provide an estimate of an individual's importance within the group. A dancer with a status of $0$ has no influence since no one else in the group is observing that dancer. A dancer with a status of $1$ has the maximum possible influence as every other individual is directly observing that dancer. However, due to the time-varying nature of the graph, an individual's importance depends not only on the current node status but also on its node status in the past. Therefore, as a first estimate of instantaneous importance we can look at each node's average status over the past $1$ second. A plot of averaged node status for part of the tracked period is shown in Figure \ref{fig:status}.

Although node status provides a measure of an individual's potential to influence the group, it does not indicate whether that influence was actually exercised. Therefore, to examine if node status is indeed related to the influence of a dancer in this data set, we investigate a quantitative measure of an individual's influence on the rest of the group. This quantity is the time, referred to as {\em lead time}, at which a peak occurred in the cross-correlation function between an individual's direction of motion and the direction of the group's motion. Positive values for this lead time indicate that an individual tended to \emph{lead} the group (i.e. change direction and then have the group follow) while negative values indicate that an individual tended to \emph{lag} the group (i.e. change direction to follow the group after the group had changed). We found that this lead time measure correlates strongly with average node status, with the nodes with highest average status having the largest lead times and the nodes with lowest average status having the largest lag times. Thus, we argue that our node status computations do indeed provide insight into leadership roles within the group.  Importantly, the ability to calculate node status at any point in time allows us to investigate instantaneous and changing leadership throughout the flocking event.

By examining our plot of averaged node status, we can identify  ``leadership events'' where one particular node achieved the greatest importance within the group (with a high status value) for an extended period of time. In Figure \ref{fig:status} we can observe one such event when node $10$ became a leader between approximately $28.75$ and $31.45$ seconds. Looking at the video, it can be observed that during this time the group was moving from the back left corner of the room toward the front right corner, with node $10$ at the front of the group. This suggests that our node status measurements can  capture emergent leadership.

Another leadership event can be observed from the data during a period when one dancer stopped moving and the remaining dancers started circling around this individual. However, the individual with the highest status during this event was not the stationary one, but rather one who was very close by the stationary one and who kept moving in a circle. This seems particularly interesting since at other times (however not during our tracked period) one dancer would stop and the whole group would eventually stop too.   The difference between these two kinds of events (circling versus stationary group motion) may be due to the differences between the status of the stationary and nearby dancers in the first case as compared to the second case.  

\begin{figure}
\centering
\includegraphics[width=\maxwidth]{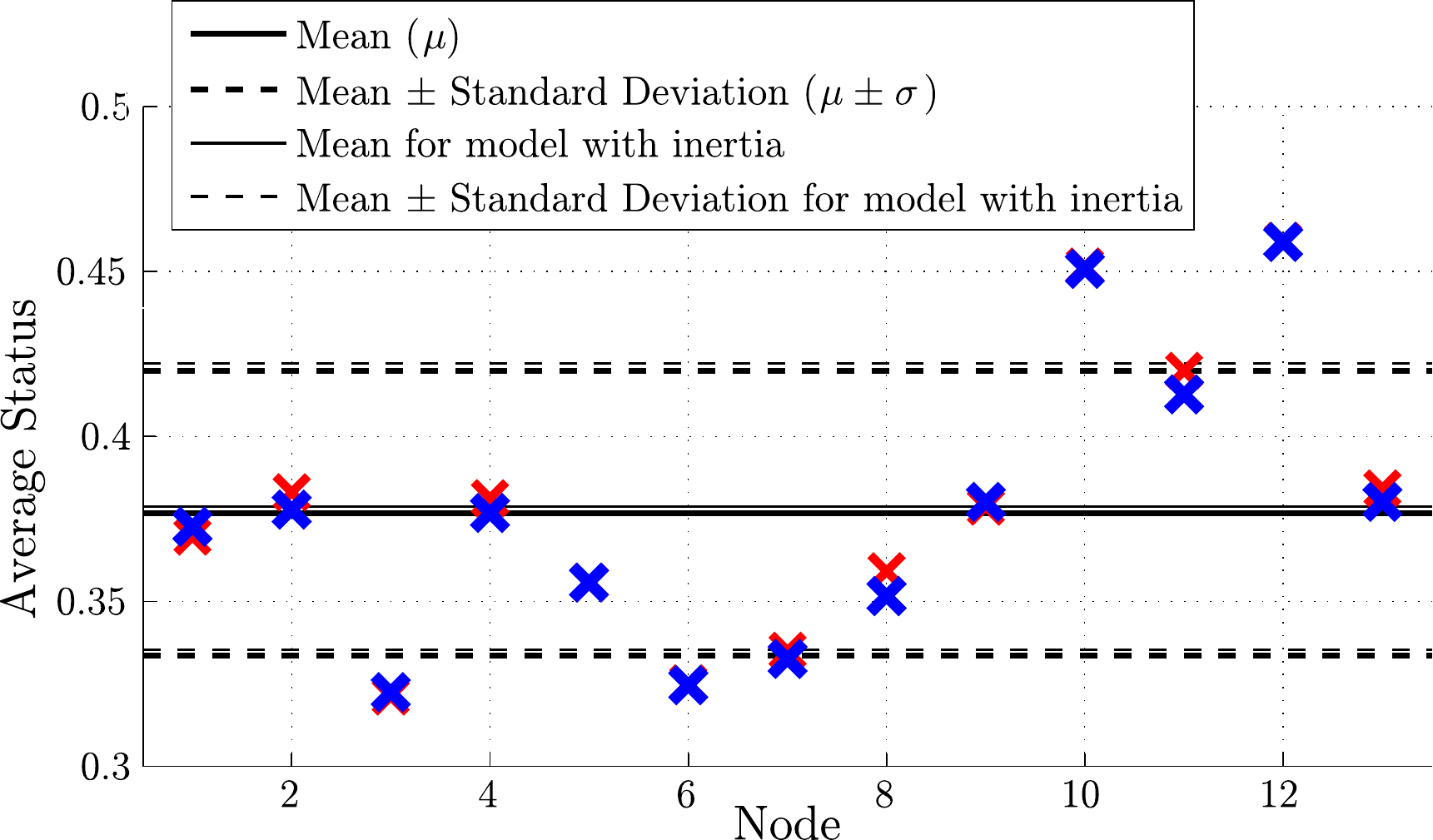}
\caption{Average node status over the whole tracked period.  The blue crosses represent the status values without any inertia term, while the red crosses represent the status values with an inertia term where the probability resets to 0 when new neighbors are chosen and grows back to 1 with an increment of 0.14 per frame.}
\label{fig:avestatus}
\end{figure}

By averaging each individual's status over the whole tracked period we can further evaluate whether some individuals had a disproportionate influence on the group.  Figure \ref{fig:avestatus} shows the average of each node's status over the tracked period. We can see that nodes $12$ and $10$ had the highest average status, with values $1.9\sigma$ and $1.7\sigma$ higher than the group mean, where $\sigma$ is the standard deviation of the average status values over all nodes.  

The  average of  an individual's status over the whole tracked period can be similarly computed in the case that we include a probability-based reluctance to switch neighbors, as described above to model switching inertia or, equivalently, commitment to neighbors.   Interestingly, while this inertia term does lead to a significant decrease in average neighbor changes per second,  the overall structure of average node status values does not change significantly. Figure \ref{fig:avestatus}  displays the average status of each node both for the original model (blue) and the filtered model with the inertia term (red). Although there is some variation between the average node status values for the two models, the same nodes represent upper and lower outliers. This suggests that the incorporation of reluctance to switch neighbors into the model does not alter the overall sensing structure, notably the emergence of leaders, even as it potentially smooths out unrealistic fluctuations of node neighbors.

We hypothesize that  the emergence of the outlier nodes  in Figure \ref{fig:avestatus}, and in particular those with very high status, is due to human bias.  To test this  we developed an agent-based flocking simulation which lacks any human bias.  Given our hypothesis, we would expect that the simulated agents, without human bias, would not exhibit outlier nodes.  The simulation models  particles that move in the plane, in a space with boundaries like in the dance studio,  and follow the  rules and parameter values close to those given to the dancers. The simulation was run in Matlab with particle positions  updated synchronously to move in the direction to maintain one arm's length (assumed to be $0.80$ m) from its two closest nodes within a viewing angle of $180^\circ$, while also repelling from all other nodes within an arm's length. Additional functions were incorporated to provide limits on velocities, turning rates and response to boundaries.  The corresponding node status of each of these particles was calculated analogously to those of the  dancers.  

\begin{figure}
\centering
\includegraphics[width=\maxwidth]{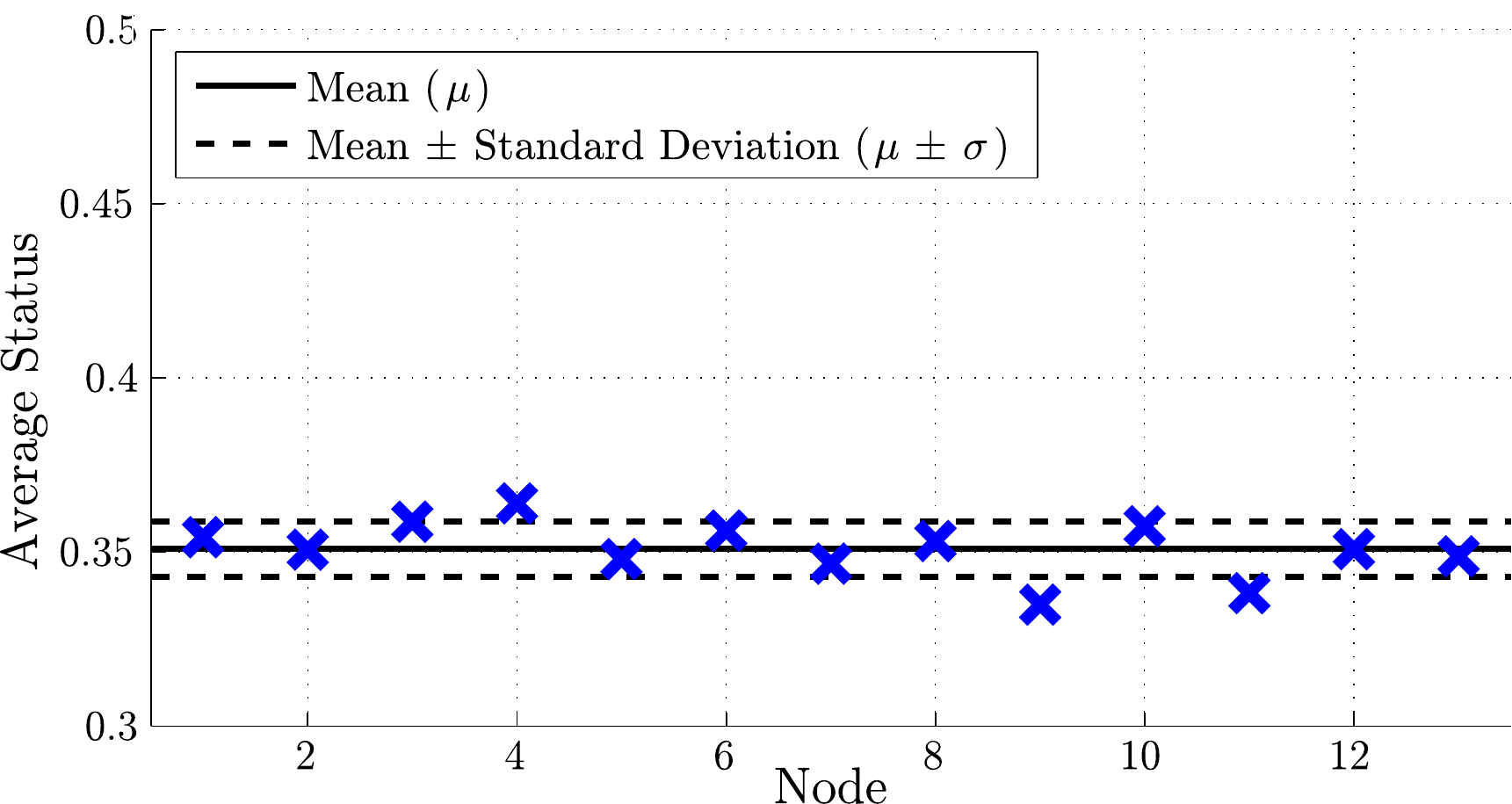}
 \caption{Average node status, with no inertia, from a simulation over an equivalent period to the tracked dancer data.}
\label{fig:statussim} 
\end{figure}

Our simulated system was initialized with positions and headings matching those of the dancers in our experiment and then the average status of each node was calculated over the following 72 seconds. The average node status over the first 72 seconds is shown in Figure \ref{fig:statussim} and can be compared to the plot in Figure \ref{fig:avestatus}.   It can be seen in Figure \ref{fig:statussim} that both the mean and the standard deviation are smaller than in the case of the human dancers and furthermore there are no significant outliers.  Every node in the simulation has a status in the range of values below the human dancer that ranked eighth in terms of highest node status and above the human dancer that ranked tenth.  Additional simulations were also run with random initial conditions. The average mean status and average standard deviation of 40 simulations run for 72 seconds each with random initial conditions was found to be approximately 0.352 and 0.0094, respectively. These mean and standard deviation values are very similar to those from the simulation of Figure \ref{fig:statussim} with the dancers' initial conditions (0.351 and 0.0080, respectively) and are significantly smaller than what is computed from the tracked dancer data (0.377 and 0.043, respectively). 

This comparison suggests that one consequence of  human bias in the behavior of the dancers was  that some individuals were less influential and other individuals significantly more influential as compared to a group of identical particles. This implies that rather than leadership simply arising as a result of random mixing within the group, the behavior of some individuals makes them more likely to assume positions of high influence.   We note that the dancers corresponding to nodes  10, 11 and 12 (the three nodes with highest average status) are three of the four dancers in the small disconnected group of Figures~\ref{fig:trackedframe} and \ref{fig:flockgraph}, suggesting  further possible consequences of emergent leaders.

\section{Final Remarks}\label{final}

The Flock Logic project, conceived at the intersection of dance and control theory,  produced a novel and generative framework for artistic, engineering and scientific investigation of collective motion.   The project centered around explorations and experiments with the motion patterns that emerge when dancers apply feedback rules modeled after those attributed to flocking birds or schooling fish.    The framework combined the prescribed rules of individual behavior and response with the unknown choices and actions of living agents, yielding  opportunities for exploration that was part systematic and part uncontrolled.     As a result, the Flock Logic framework  proved useful for artistic exploration of dance and tools for choreography, for engineering exploration of decentralized control laws for multi-agent system dynamics, as well as for scientific investigation of collective animal behavior and crowd dynamics.

The explorations built off of a set of basic ``flocking'' rules of cohesion and repulsion:  dancers were instructed to move around while maintaining an arm's length distance from a prescribed number of other dancers and not letting anyone come closer than arm's length.   Rules for walls, obstacles, and zones were added.    Additional objectives and preferences were imposed selectively and secretly so that a small subset of dancers were asked to influence the group only through their motion and without explicit signaling.   Dancers behaving as pursuers or predators applied pressure dynamically on the group, often to beautiful effect.   Synchrony and anti-synchrony of directionality were explored using alignment rules.    Many artistic explorations made use of different kinds of rules that were originally motivated from observations or analysis of animal behavior or from engineering design objectives, such as foraging and coverage.  Other explorations were motivated purely by artistic goals, such as designed responses to specific obstacles.

To address questions concerning the role of the heterogeneity of the group of dancers and specifically the relative influence of the different dancers on the collective motion, we analyzed video data of an experiment with thirteen dancers applying the basic rules of flocking with two-person cohesion.   From the video we tracked the trajectories of the dancers over a 72-second segment.   Then, we applied the flocking rules to the data to estimate the network graph at each frame of the video, that is, who was paying attention to whom at each time step.   From the resulting time-varying graph we computed  node status for each dancer at each frame; node status provides a measure of how much attention a dancer received from the rest of the dancers.   We discussed how node status was strongly correlated with lead time in turning, i.e., dancers with high status would typically turn before the rest of the group.  From this we argued that high node status suggests high influence and therefore leadership. By examining the average status of each dancer over the whole tracked segment, we found  two of the dancers with status higher than the mean value by nearly twice the standard deviation.   We showed how this result is robust to the addition of an inertia term that models a dancer's commitment to its newly acquired neighbors.   We also showed evidence that human bias explains the large variation in influence among the dancers, and in particular the outliers, by comparing the data with analogous results from a simulation of dancers modeled as particles without human bias. 

These results raise many more interesting questions and possibilities for future investigation.   For example, how does human bias produce leadership, where no such leadership was assigned?   Do certain dancers move in ways that attract the attention of the others?   Or do individuals who emerge as leaders break the rules, for example, by paying less attention to others than instructed?   The results suggest the possibility of an interesting tension between following rules and breaking rules.  This could be explored scientifically using evolutionary game theory in which there is a benefit to breaking the rules associated with influencing the group toward one's own preferences but also a cost to breaking the rules associated with losing the advantages of group living.

Other questions concern the relationship between the rules and environment and the resulting shape and momentum of the group.  What accounts for polarized versus circular motion?  What accounts for fissions and fusions of the group?  Many further artistic, engineering and scientific investigations are possible, even extending the basic flocking rules into more abstract, non-spatial, domains.   Human flocking for recreation or therapy might also be explored -- participants in the Flock Logic performances described finding it calming to engage with a group without a goal and rewarding to be part of creating something new.

\section{Acknowledgments}

The  authors thank Alex Holness for his contributions to the study of the lead/lag time and its correlation with node status.




\end{document}